# $Ar^+$-ion Bombardment of $TiO_2$ Nanotubes Creates Co-catalytic Effect for Photocatalytic Open Circuit Hydrogen Evolution

Xuemei Zhou[a], Ning Liu[a], Patrik Schmuki[a,b]*

[a]Department of Materials Science WW4, LKO, University of Erlangen-Nuremberg, Martensstrasse 7, 91058 Erlangen, Germany.

[b]Department of Chemistry, Faculty of Science, King Abdulaziz University, P.O. Box 80203, Jeddah 21569, Saudi Arabia

*Corresponding author. Tel.: +49 91318517575, Fax: +49 9131 852 7582

Email: schmuki@ww.uni-erlangen.de

**Abstract**

In the present work, extended Argon ion bombardment was used to modify anatase $TiO_2$ nanotube (NTs) layers. In situ X-ray photoelectron spectroscopy revealed a high content of sub-oxide formed ($Ti^{3+}$, $Ti^{2+}$ states) in the nanotube layers. These samples were tested for their ability to show open-circuit photocatalytic hydrogen evolution. We found for UV (cw laser 325nm) illumination as well as for AM1.5 conditions, a strongly enhanced $H_2$ evolution activity in absence of the usual noble metal-nanoparticle decoration. We conclude therefore that these sub-oxide states can play an important role in the activation of $TiO_2$ nanotubes - namely as a $H_2$ evolution co-catalyst.

**Introduction**

In 1971, Fujishima and Honda reported on the use of a $TiO_2$ for triggering photocatalytic reactions, i.e. by shining light of an energy higher than the band gap of the semiconductive $TiO_2$ (3.2 eV for anatase or 3.0 eV for rutile) electron-hole pairs are generated that then may trigger red-ox processes with the environment [1]. Due to the relative position of $TiO_2$ valence and conduction band with respect to important red-ox potentials (namely water), a wide range of useful red-ox processes are thermodynamically feasible [2,3]. This may involve hole transfer for oxidative photocatalytic destruction of pollutants or electron transfer to generate $H_2$ from $H_2O$ [4,5]. In general, photocatalytic reactions are preferably carried out on large specific surface area $TiO_2$ entities such as nanoparticle suspensions [6,7] or nanostructured electrodes [8-10]. Most recently, particularly self-organized $TiO_2$ nanotube arrays have received considerable attention for photocatalysis because of their unique combination of geometry with favorable charge-separation properties [11-15].

One of the most desired photocatalytic reactions is the generation of $H_2$ and $O_2$ from water. The reaction is generally however very sluggish on the bare $TiO_2$ surface, and therefore adequate co-catalysts (noble-metals such as Au, Pt, Pd [16-20]) are needed to achieve considerable reaction rates. Recently, it was however reported that anatase $TiO_2$ nanotube layers after a treatment in $H_2$ under high pressure conditions are activated to provide considerable open circuit hydrogen evolution rates, i.e. hydrogenation seems to be able to provide specific intrinsic co-catalytic activation of anatase nanotubes and thus enable considerable co-catalyst free $H_2$ evolution rates [21]. The work further investigated also common reduction treatments [17,22-25] of $TiO_2$ such as Ar, or Ar/$H_2$ annealing under elevated temperatures, but these treatments were found not to induce any co-catalytic effect [21]. In the present work we show, however that $TiO_2$ nanotube layers that are exposed to $Ar^+$ -ion sputtering show a considerably enhanced hydrogen production rate that can directly be attributed to ion-beam generated sub-oxide states ($Ti^{3+}/Ti^{2+}$).

**Experimental section**

TiO$_2$ nanotubes (NTs) were grown by anodization of titanium foils (0.125mm thickness, Advent, purity 99.6+%) in ethylene glycol (EG, Sigma–Aldrich), water (1M) and NH$_4$F (0.2M, Sigma–Aldrich, 98%) electrolyte at 60V (power supply, VOLTRAFT VLP2403Pro) for 20 min in a two electrode configuration using a platinum counter electrode. Before anodization, the Ti foils were degreased in acetone and ethanol by sonication, respectively, rinsed with deionized water, and dried in nitrogen atmosphere. In order to convert the amorphous tube layer to anatase, thermal annealing was performed in air using rapid thermal annealer (Jipelec JetFirst 100) at 450 ℃ for 1h with a temperature ramp rate of 0.5 ℃/s.

X-ray photoelectron spectrometer (XPS, PHI 5600 XPS spectrometer, US) was used for ion bombardment, Ar$^+$-ion sputtering and compositional analysis at the sample. For bombardment, the crystallized TiO$_2$ NTs were exposed to argon ion beam sputtering (3.5kV, current density 15mA). The sputtered sample area was 5×5mm, sputtering was carried out using a sequence of 10min sputtering followed by 30s spectra acquisition. The incident angle of Argon beam on the surface of TiO$_2$ NTs is 36$^o$ – this corresponds to a nominal sputter rate of 0.85nm/min on a SiO$_2$ reference layer. XPS spectra were acquired using monochromatic X-rays with a pass energy of 23.5eV. All the XPS element peaks are shifted to the Ti2p standard position.

For morphological characterization of the TiO$_2$ NT layers, we used a field emission scanning electron microscope (Hitachi S4800). XRD patterns of the crystallized TiO$_2$ NTs were collected using an X-ray diffractometer (X'pert Philips PMD diffractometer) with a Panalytical X'celerator detector and CuKa radiation (λ = 1.54056Å).

For open circuit photocatalytic H$_2$ evolution measurements we immersed the TiO$_2$ NT layers in an aqueous ethanol (20 vol%) solution and illuminated the surface either with a 325nm cw-HeCd laser (60 mW/cm$^2$, 200mW, Kimmon, Japan) or an AM 1.5 (100 mW/cm$^2$) solar simulator for 12h, respectively. A gas chromatograph (GCMS-QO2010SE, SHIMADZU) with TCD detector was used to obtain the amount of generated H$_2$ for different TiO$_2$ NTs samples.

**Results and discussion**

In order to generate and quantify $Ar^+$-ion bombardment effects (namely inducing the reduction of $TiO_2$ to $Ti^{3+}$), we carried out these steps in the UHV of an XPS. Ti sub-oxides are reported to be very sensitive to environmental exposure [26,27], this guaranteed an identical modified starting point for further investigations. The samples were then transferred to photocatalytic $H_2$ evolution measurements. After these measurements the sample sub-oxide status was then reassessed by XPS and a number of reference experiments (such as exposing $Ar^+$ bombarded samples to ambient) were carried out.

Fig 1a shows SEM images of the nanotube layers used in this work after anodic growth in a $NH_4F/H_2O$/ethylene glycol electrolyte and after the samples were air annealed at 450°C to transform the samples to anatase. From Fig. 1a and b it is apparent that the tubes have a diameter of 85 nm in average and a length (layer thickness of 6.6µm). After $Ar^+$-ion sputtering for 5h, as described in the experimental section, the tube morphology at the top of the layer shows typical $Ar^+$-ion sputtering marks (Figure 1c). The sputter conditions used correspond to a nominal removal of a compact $TiO_2$ layer of 255 nm from the top. As apparent from the cross section SEM, the layer thickness was indeed diminished by the sputter process only by 0.24 µm (Figure 1b and d). XRD in Figure 1e shows that the overall anatase structure of the $TiO_2$ nanotubes remained after 5h $Ar^+$-ion bombardment.

From XPS spectra that were acquired before (Fig.2a) and after $Ar^+$-ion bombardment (Fig. 2b-d), clear alterations in the Ti2p peak as well as in the O1s peak are apparent after sputtering. By fitting of the Ti2p peaks according to literature [28], the individual peaks can be ascribed to $Ti^{4+}$ at 463.6eV ($Ti2p_{1/2}$) and 457.9eV ($Ti2p_{3/2}$), $Ti^{3+}$ at 461.8eV ($Ti2p_{1/2}$) and 456.1eV ($Ti2p_{3/2}$), $Ti^{2+}$ at 459.9eV ($Ti2p_{1/2}$) and 454.1eV ($Ti2p_{3/2}$). Compared to Ti2p before ion bombardment, an obvious increase of sub-oxide peaks is evident from Figure 2b and 2c. For $TiO_2$ NTs after 50min sputtering, the measured contents are 43.85% for $Ti^{4+}$, 32.23% for $Ti^{3+}$ and 23.92% for $Ti^{2+}$, for $TiO_2$ NTs after 5h sputtering, $Ti^{2+}$ rises to 41.77%, $Ti^{3+}$ to 29.30%, and $Ti^{4+}$ drops to 28.93%, respectively. Also a significant alteration of the O1s peak can be observed with a shift in the maximum from 529.1eV to 529.9eV and strong contributions at higher binding energies (Figure 2d), which is well in line with the formation of sub-oxides in the $TiO_2$ NTs.

Surface $Ti^{3+}$ is susceptible to air exposure as evident from XPS spectra taken after the samples were removed from the XPS environment and reintroduced after being exposed to air (for 2 h or 7 days). Clearly exposure to air for 2 h leads to a drastic re-oxidation of the samples: the measured $Ti^{2+}$ content decreases from 41.77% to 7.79%, the $Ti^{3+}$ content from 29.30% to 7.66% (Figure 2e), and accordingly the partial re-oxidation is also apparent from a shift of O1s peak from 530.1eV to 529.4eV. After exposure for 7 days in air, the sub-oxide species reach a 1% level, i.e. as shown in inset of Figure 2e. Quantitative experimental data are compiled (Table 1f) in Fig.1.

Figure 3 shows the observed hydrogen evolution rate of the bare $TiO_2$ NTs and after they have been exposed to $Ar^+$ sputtering measured under open circuit conditions in a 20 vol% ethanol/DI water solution. For the sputtered samples, clearly more $H_2$ is generated both under UV laser and AM1.5 illumination, i.e. for the $Ar^+$ bombardment $TiO_2$ NTs under plain UV illumination, the amount of $H_2$ is 2-fold that of the bare $TiO_2$ NTs after the same time illumination. For AM1.5 illumination, for the non-treated tubes there is no significant $H_2$ evolution detectable, whereas clearly the sub-oxide containing sample shows obvious $H_2$ evolution.

In order to assess how much the results are affected by environmental deterioration of the sub-oxide content, we additionally investigated samples that were exposed to air for 2 h and 7 days (Figs.2e and f) for their $H_2$ evolution performance (Fig.3b). Clearly the sample exposed for 2 h to air still shows a $H_2$ evolution activity comparable to a fresh sample, while for the sample exposed for 7 days to air, the activity has virtually returned to the untreated $TiO_2$ nanotube samples. This clearly shows that the $H_2$ evolution photocatalytic activity is related to the presence of sub-oxide states ($Ti^{3+}/Ti^{2+}$). However, it is very interesting to note that the sample after 2 h air exposure shows a drastic drop in XPS-accessible $Ti^{3+/2+}$ (from 29.30 % to 7.66% for $Ti^{3+}$, and 41.77% to 7.79% for $Ti^{2+}$, respectively), while the $H_2$ activity remains (within the error of detection) almost unaffected. This may indicate that $Ti^{3+/2+}$ also in much lower concentrations than observed directly after sputtering provide sufficient co-catalytic centers to be not rate determining. On the other hand, it may indicate that $Ti^{3+}$ centers that are buried [i.e., are not directly on the surface (XPS penetration depth)] may contribute to activity. Nevertheless, if re-oxidation of the samples is too extended they may be buried too deep to be effective.

In the light of recently reported effects of a high temperature hydrogenation of $TiO_2$

NTs on the creation of a *stable* co-catalytic center for photocatalytic open circuit $H_2$ evolution [21], the present work allows same conclusions. It is shown that sub-oxide ($Ti^{3+/2+}$) states can represent a co-catalytic center for $H_2$ evolution and may thus constitute a partial explanation for such findings. Nevertheless, in contrast to hydrogenation, the co-catalytic effect produced by $Ar^+$-ion bombardment is not stable, i.e. it fades out by environmental interactions.

Overall it can thus be concluded that Ti-sub-oxide states can represent co-catalytic centers that allow an enhanced photocatalytic hydrogen evolution from $TiO_2$ nanotubes. In order to make the co-catalytic effect "permanent", a passivation of these states is required as it is possibly provided by hydrogenation.

**Conclusions**

The present work shows that $Ar^+$-ion bombardment creates sub-oxide states in the oxide of $TiO_2$ nanotubes that beneficially affect the photocatalytic hydrogen evolution rate under open circuit conditions. The finding thus may open new perspectives of replacing noble metal co-catalysts in photocatalytic water splitting.

**Acknowledgement**

The authors would like to acknowledge ERC, DFG and the Erlangen DFG cluster of excellence for financial support.

**Figures**

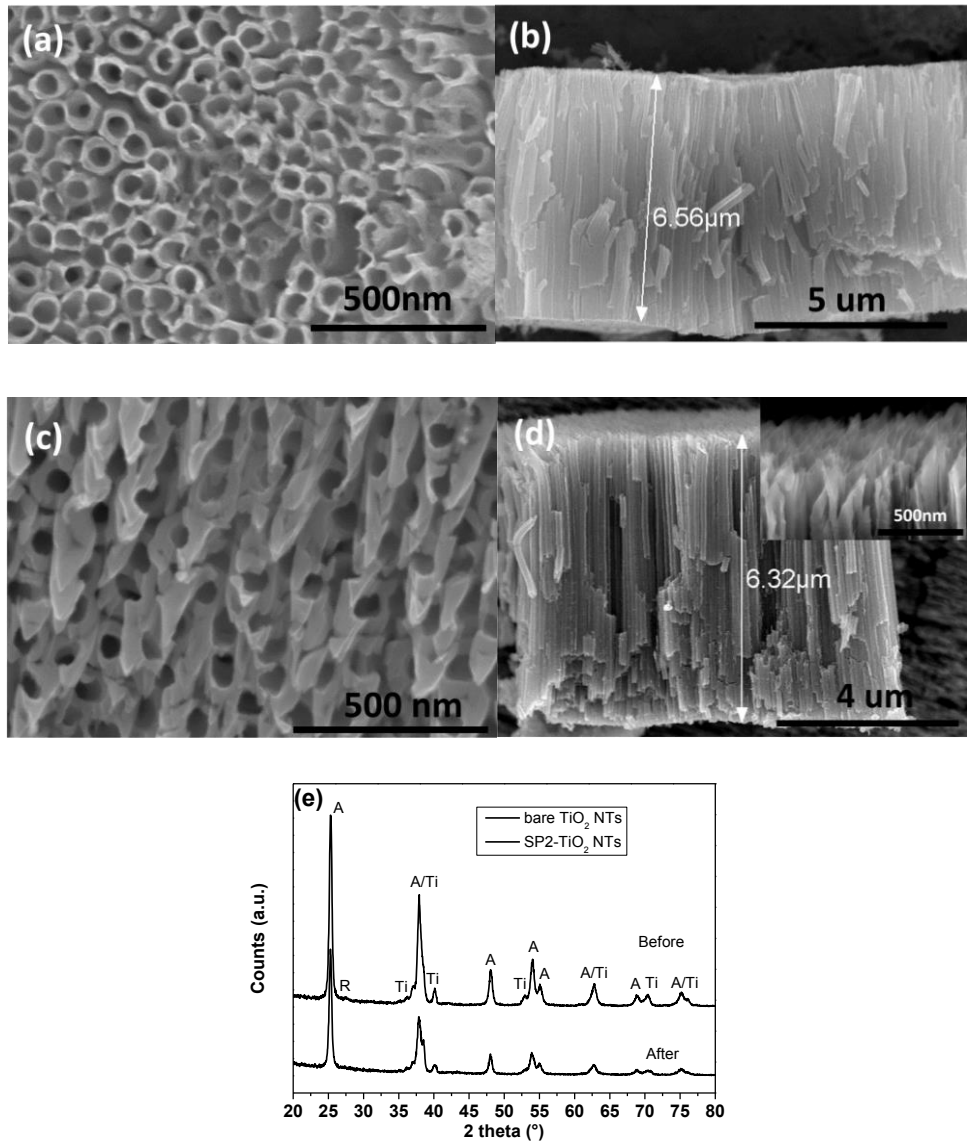

(f)

| Samples | Ti/O | $Ti^{4+}$ | $Ti^{3+}$ | $Ti^{2+}$ |
|---|---|---|---|---|
| $TiO_2$ NTs | 0.465 | 100 | 0 | 0 |
| SP1-$TiO_2$ NTs[*1] | 0.502 | 43.85 | 32.23 | 23.92 |
| SP2-$TiO_2$ NTs[*2] | 0.528 | 28.93 | 29.30 | 41.77 |
| SP2-$TiO_2$ NTs-Air 2h | 0.473 | 84.55 | 7.66 | 7.79 |
| SP2-$TiO_2$ NTs-Air 7d | 0.431 | 95.49 | 1.37 | 3.14 |

**Figure 1.** Top (a) and side view (b) of SEM images of $TiO_2$ NTs before and after (c), (d) 5h ion bombardment. (e) XRD patterns of annealed $TiO_2$ NTs before and after 5h ion bombardment. (f) Table with quantitative compositional data calculated from XPS peak fitting. [*1]corresponds to a sputtering time of 50min. [*2]corresponds to a sputtering time of 5h.

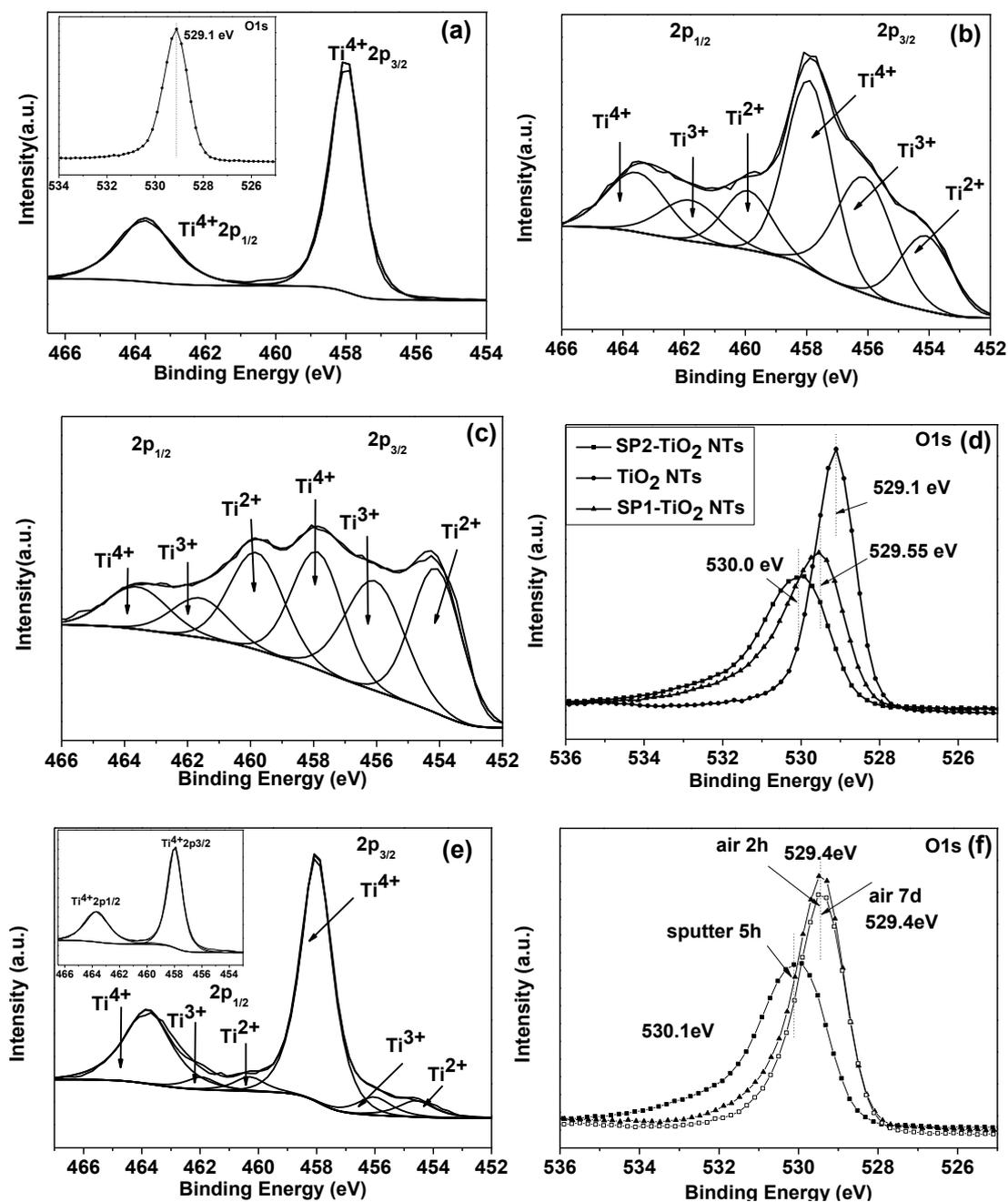

**Figure 2.** High resolution Ti2p peaks before (a) and after 50min (b), 5h (c) Argon bombardment, high resolution O1s peaks (d). Inset of Figure 2a is the high resolution O1s peak for TiO$_2$ NTs. High resolution Ti 2p (e) and O1s (f) for 5h sputtered TiO$_2$ NTs after 2h and 7 days air exposure. Inset of Figure 2e is the high resolution Ti 2p for 5h sputtered TiO$_2$ NTs after 7days air exposure.

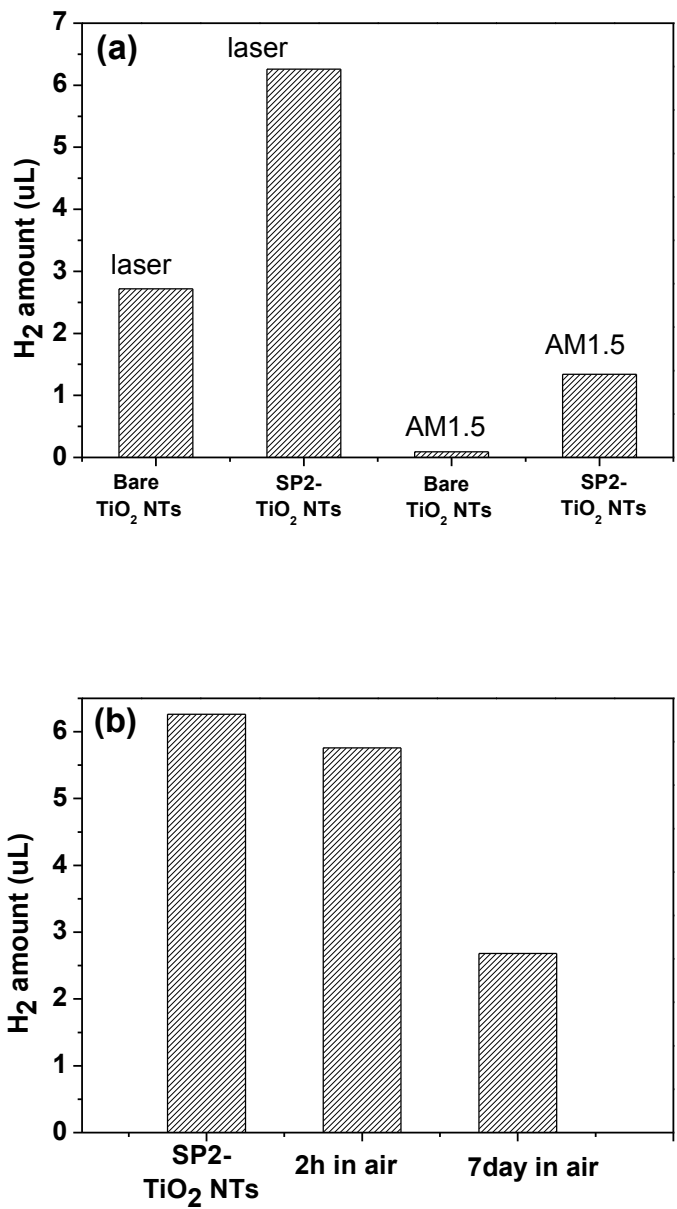

**Figure 3.** (a) Open circuit hydrogen generation for TiO$_2$ NTs after Argon bombardment and (b) after exposure to air for certain time.